# Methods for calculation of the coupling coefficients in the Coupling Cavity Model of arbitrary chain of resonators

M.I. Ayzatsky[1], V.V.Mytrochenko

National Science Center Kharkov Institute of Physics and Technology (NSC KIPT),
610108, Kharkov, Ukraine

We present the short description of the methods for calculation of the coupling coefficients in the Coupling Cavity Model of arbitrary chain of resonators. In the first part the procedure that is based on the Mode Matching Method is given. Then we present the new method that used only one eigen vector.

### 1 Introduction

Development of the new Coupling Cavity Model (CCM) on the base of the Mode Matching Method (MMM) gave possibility to look more deeply into the properties of the inhomogeneous disk loaded waveguides (DLWs) [1,2,3,4,5], especially on the basis of methods that are used for tuning such DLWs [6,7]

In the frame of the CCM such coupling equations can be obtained [1-5]

$$Z_k e_{q_0}^{(k)} = \sum_l e_{q_0}^{(l)} \alpha_{q_0}^{(k,l)},$$

where $e_{q_0}^{(k)}$ - amplitudes of the basis $E_{q_0}$ mode, $Z_k = 1 - \omega^2/\omega_{q_0}^{(k)2}$, $\omega_{q_0}^{(k)}$ - eigen frequency of this mode in the k-cell, $\alpha_{q_0}^{(k,l)}$ - coupling coefficients that depend on the frequency and the geometric sizes of coupling volumes.

There are some drawbacks in the developed approach for calculating the coupling coefficients $\alpha_{q_0}^{(k,l)}$. First one is the possibility of conducting full numerical simulation of nonuniform DLW and obtaining all necessary coupling coefficients for the simple cell geometry only. The CCM, that was developed recently [3-5], can be used only for geometries for which there are analytical expressions of the eigen functions. Secondly, for receiving the necessary accuracy of simulation we must take into account the great number of eigen functions that can be difficult from several reasons (accuracy of Bessel function calculation, etc).

Therefore, it is necessary to develop the general approach that gives possibility to obtain the coupling coefficients for arbitrary chain of resonators without using the great number of eigen functions.

## 1 Procedures for calculation of the coupling coefficients

We now consider the general problem obtaining the coupling cavity equations for arbitrary waveguides. We present the short description of the procedure that is based on the field expansion in terms of eigen vectors. Then we describe the new procedure that used only one eigen vector. Short description of this procedure is presented in [7].

### 1.1 Procedure for calculation of the coupling coefficients on the base of the Mode Matching Method

We will assume that all field quantities have a time variation given by $\exp(-i\omega t)$. The electrical charge and current are assumed to be absent in the waveguide. The electromagnetic field is governed by the Maxwell's equations

---

[1] M.I. Aizatskyi, N.I.Aizatsky; aizatsky@kipt.kharkov.ua

$$\nabla \times \vec{E} = i\omega\mu_0\vec{H},$$
$$\nabla \times \vec{H} = -i\omega\varepsilon_0\varepsilon\vec{E}. \quad (1.1)$$

Let's divide the waveguide into the set of arbitrary volumes $V_k$ as shown in Fig. 1 and represent electromagnetic field in each volume $V_k$ as[2]

$$\vec{E} = \sum_q e_q^{(k)} \vec{\mathbb{E}}_q^{(k)}(\vec{r}) \quad (1.2)$$

$$\vec{H} = i\sum_q h_q^{(k)} \vec{\mathbb{H}}_q^{(k)}(\vec{r}), \ \vec{r} \in V_k, \quad (1.3)$$

where $\vec{\mathbb{E}}_q^{(k)}, \vec{\mathbb{H}}_q^{(k)}$ - solenoidal eigen vectors for the volume $V_k$, $\omega_q^{(k)}$ - eigen frequencies for the volume $V_k$. Vectors $\vec{\mathbb{E}}_q^{(k)}, \vec{\mathbb{H}}_q^{(k)}$ satisfy such equations

$$\nabla \times \vec{\mathbb{E}}_q^{(k)} = i\omega_q^{(k)}\mu_0\vec{\mathbb{H}}_q^{(k)},$$
$$\nabla \times \vec{\mathbb{H}}_q^{(k)} = -i\omega_q^{(k)}\varepsilon_0\varepsilon\vec{\mathbb{E}}_q^{(k)}, \ \vec{r} \in V_k \quad (1.4)$$

with boundary conditions $\vec{\mathbb{E}}_{q,\tau}^{(k)} = 0$ at the volume $V_k$ border.

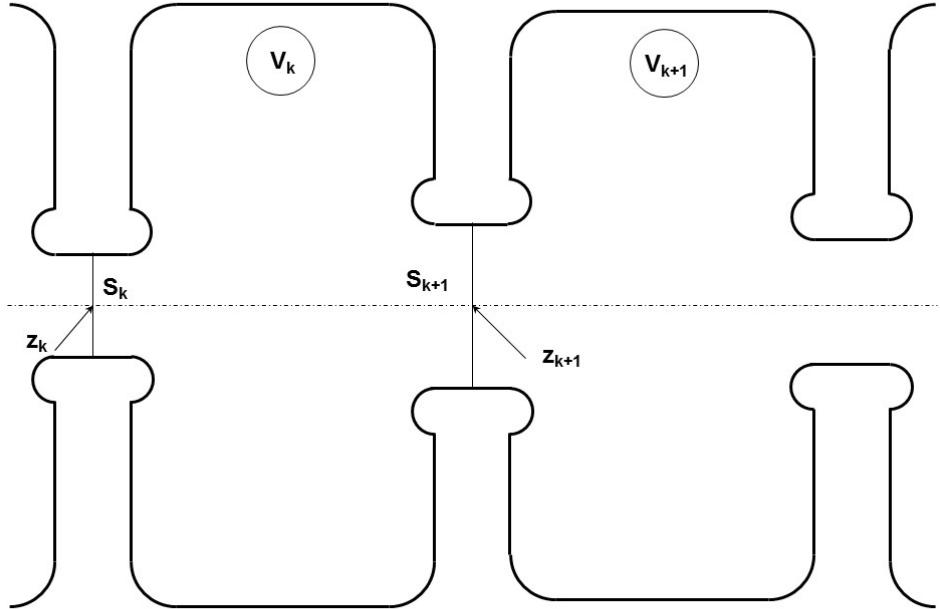

Fig. 1

Amplitudes $e_q^{(k)}$ and $h_q^{(k)}$ can be expressed through the tangential components of the electric fields on the surfaces $S_k$ and $S_{k+1}$

$$\left(\omega_q^{(k)2} - \omega^2\right)e_q^{(k)} = \frac{i\omega_q^{(k)*}}{N_q^{(k)}}\frac{|\varepsilon|}{\varepsilon}\left(\oint_{S_k}[\vec{E}_\tau^{(k)}\vec{\mathbb{H}}_q^{(k)*}]d\vec{S} + \oint_{S_{k+1}}[\vec{E}_\tau^{(k+1)}\vec{\mathbb{H}}_q^{(k)*}]d\vec{S}\right), \quad (1.5)$$

$$h_q^{(k)} = -i\frac{\omega}{\omega_q^{(k)*}}\frac{\varepsilon}{|\varepsilon|}e_q^{(k)}, \quad (1.6)$$

where

---
[2] We will consider electromagnetic fields with amplitudes of potential modes equal zero

$$N_{m,n}^{(k)} = \mu_0 \int_{V_k} \vec{\mathbb{H}}_q^{(k)} \vec{\mathbb{H}}_q^{(k)*} dV = \varepsilon_0 |\varepsilon| \int_{V_k} \vec{\mathbb{E}}_q^{(k)} \vec{\mathbb{E}}_q^{(k)*} dV, \qquad (1.7)$$

Index $\tau$ here and bellow designates the tangential components of vector

$$\vec{E}_\tau = \left(\vec{e}_x E_x + \vec{e}_y E_y + \vec{e}_z E_z\right)_\tau = \left(\vec{e}_x E_x + \vec{e}_y E_y\right) \qquad (1.8)$$

There is a complete orthogonal set of vectors $\{\vec{f}_{\tau,s}^{(k)}(\vec{r}_\perp)\}$ (the basis vectors) in the transverse aperture $S_k$ and we can decompose $\vec{E}_\tau^{(k)}(\vec{r}_\perp)$ as

$$\vec{E}_\tau^{(k)} = \sum_s C_s^{(k)} \vec{f}_s^{(k)}(\vec{r}_\perp) \qquad (1.9)$$

Combining (1.6) and (1.9), we arrive at the expression

$$h_q^{(k)} = -i\omega \sum_s \left(C_s^{(k)} F_{q,s}^{(k,k)} + C_s^{(k+1)} F_{q,s}^{(k+1,k)}\right) \qquad (1.10)$$

where

$$F_{q,s}^{(k,k)} = \frac{i}{\left(\omega_q^{(k)2} - \omega^2\right) N_q^{(k)}} \oint_{S_k} [\vec{f}_{\tau,s}^{(k)} \vec{\mathbb{H}}_q^{(k)*}] d\vec{S}$$

$$F_{q,s}^{(k+1,k)} = \frac{i}{\left(\omega_q^{(k)2} - \omega^2\right) N_q^{(k)}} \oint_{S_{k+1}} [\vec{f}_{\tau,s}^{(k+1)} \vec{\mathbb{H}}_q^{(k)*}] d\vec{S} \qquad (1.11)$$

Matching of the fields $\vec{H}_\tau$ [3] at the interface $z = z_k$ leads to the following relations

$$\vec{H}_\tau\left(\vec{r}_\perp, z_{k+1} - 0\right) = i\sum_q h_q^{(k)} \vec{\mathbb{H}}_{q,\tau}^{(k)}\left(\vec{r}_\perp, d_{k-1}\right) = \vec{H}_\tau\left(\vec{r}_\perp, z_{k+1} + 0\right) = i\sum_q h_q^{(k+1)} \vec{\mathbb{H}}_{q,\tau}^{(k+1)}\left(\vec{r}_\perp, 0\right) \quad (1.12)$$

Decomposing of the left and right hand parts of the equality(1.12) in terms of a complete orthogonal set of vectors $\{\vec{f}_{\tau,s'}^{\prime(k)}(\vec{r}_\perp)\}$ (the testing vectors), we arrive at the result

$$\sum_q h_q^{(k)} T_{q,s'}^{(k,-)} = \sum_q h_q^{(k+1)} T_{q,s'}^{(k,+)} \qquad (1.13)$$

where

$$T_{q,s'}^{(k,-)} = i \int_{S_k} d\vec{r}_\perp \vec{\mathbb{H}}_{q,\tau}^{(k)}\left(\vec{r}_\perp, d_k\right) \vec{f}_{\tau,s'}^{\prime(k)}(\vec{r}_\perp)$$

$$T_{q,s'}^{(k,+)} = i \int_{S_k} d\vec{r}_\perp \vec{\mathbb{H}}_{q,\tau}^{(k+1)}\left(\vec{r}_\perp, 0\right) \vec{f}_{\tau,s'}^{\prime(k)}(\vec{r}_\perp) \qquad (1.14)$$

Traditional next step in the studding of waveguide properties is substituting (1.10) into (1.13). When this is accomplished, we obtain the coupling equations for $C_s^{(k)}$

$$\sum_s \left(C_s^{(k)} \Lambda_{s,s'}^{(k,1)} + C_s^{(k+1)} \Lambda_{s,s'}^{(k+1,2)} + C_s^{(k+2)} \Lambda_{s,s'}^{(k+2,3)}\right) = 0 \qquad (1.15)$$

where

$$\Lambda_{s,s'}^{(k,1)} = \sum_q F_{q,s}^{(k,k)} T_{q,s'}^{(k,-)}$$

$$\Lambda_{s,s'}^{(k+1,2)} = \sum_q F_{q,s}^{(k+1,k)} T_{q,s'}^{(k,-)} - \sum_q F_{q,s}^{(k+1,k+1)} T_{q,s'}^{(k,+)} \qquad (1.16)$$

$$\Lambda_{s,s'}^{(k+2,3)} = -\sum_q F_{q,s}^{(k+2,k)} T_{q,s'}^{(k,+)}$$

In the CCM we rewrite (1.12)

$$\sum_{q \neq q_0} h_q^{(k)} T_{q,s'}^{(k,-)} - \sum_{q \neq q_0} h_q^{(k+1)} T_{q,s'}^{(k,+)} = h_{q_0}^{(k+1)} T_{q_0,s'}^{(k,+)} - h_{q_0}^{(k)} T_{q_0,s'}^{(k,-)} \qquad (1.17)$$

where $q_0$ is the index of a basis mode.

---

[3] We cannot match electric field as $\vec{\mathbb{E}}_{q,\tau}^{(k)} = 0$ at the volume $V_k$ border

From (1.17) we obtain such equations

$$\sum_s \left( C_s^{(k)} \tilde{\Lambda}_{s,s'}^{(k,1)} + C_s^{(k+1)} \tilde{\Lambda}_{s,s'}^{(k+1,2)} + C_s^{(k+2)} \tilde{\Lambda}_{s,s'}^{(k+2,3)} \right) =$$
$$= ih_{q_0}^{(k+1)} T_{q_0,s'}^{(k,+)} \omega^{-1} - ih_{q_0}^{(k)} T_{q_0,s'}^{(k,-)} \omega^{-1} = \frac{e_{q_0}^{(k+1)} T_{q_0,s'}^{(k,+)}}{\omega_{q_0}^{(k+1)}} - \frac{e_{q_0}^{(k)} T_{q_0,s'}^{(k,-)}}{\omega_{q_0}^{(k)}} \quad (1.18)$$

with

$$\tilde{\Lambda}_{s,s'}^{(k,1)} = \sum_{q \neq q_0} F_{q,s}^{(k,k)} T_{q,s'}^{(k,-)}$$
$$\tilde{\Lambda}_{s,s'}^{(k+1,2)} = \sum_{q \neq q_0} F_{q,s}^{(k+1,k)} T_{q,s'}^{(k,-)} - \sum_{q \neq q_0} F_{q,s}^{(k+1,k+1)} T_{q,s'}^{(k,+)} \quad (1.19)$$
$$\tilde{\Lambda}_{s,s'}^{(k+2,3)} = -\sum_{q \neq q_0} F_{q,s}^{(k+2,k)} T_{q,s'}^{(k,+)}$$

If we find solutions of such coupling systems of linear equations

$$\sum_s \left( C_s^{+(k,l)} \tilde{\Lambda}_{s,s'}^{(k,1)} + C_s^{+(k+1,l)} \tilde{\Lambda}_{s,s'}^{(k+1,2)} + C_s^{+(k+2,l)} \tilde{\Lambda}_{s,s'}^{(k+2,3)} \right) = \frac{T_{q_0,s'}^{(k,+)}}{\omega_{q_0}^{(k+1)}} \delta_{k+1,l}$$
$$\sum_s \left( C_s^{-(k,l)} \tilde{\Lambda}_{s,s'}^{(k,1)} + C_s^{-(k+1,l)} \tilde{\Lambda}_{s,s'}^{(k+1,2)} + C_s^{-(k+2,l)} \tilde{\Lambda}_{s,s'}^{(k+2,3)} \right) = \frac{T_{q_0,s'}^{(k,-)}}{\omega_{q_0}^{(k)}} \delta_{k,l} \quad (1.20)$$

we can write the expansion coefficients $C_s^{(k)}$ as

$$C_s^{(k)} = \sum_l \left( e_{q_0}^{(l+1)} C_s^{+(k,l)} - e_{q_0}^{(l)} C_s^{-(k,l)} \right) = \sum_l e_{q_0}^{(l)} \left( C_s^{+(k,l-1)} - C_s^{-(k,l)} \right). \quad (1.21)$$

In (1.20) $\delta_{k,l}$ is the Kronecker symbol.

Equations for the amplitudes of the basis mode take then the form

$$\left( \omega_{q_0}^{(k)2} - \omega^2 \right) e_{q_0}^{(k)} = \omega_{q_0}^{(k)2} \sum_l e_{q_0}^{(l)} \alpha_{q_0}^{(k,l)}, \quad (1.22)$$

where the coupling coefficients $\alpha_{q_0}^{(k,l)}$ can be calculated with using the solutions of systems (1.20)

$$\alpha_{q_0}^{(k,l)} = \frac{i\omega_{q_0}^{(k)*}}{\omega_{q_0}^{(k)2} N_q^{(k)}} \frac{|\varepsilon|}{\varepsilon} \times$$
$$\times \left( \sum_s \left( C_{s,}^{+(k,l-1)} - C_s^{-(k,l)} \right) \oint_{S_k} [\vec{f}_{\tau,s}^{(k)} \vec{\mathbb{H}}_{q_0}^{(k)*}] d\vec{S} + \sum_s \left( C_s^{+(k+1,l-1)} - C_s^{-(k+1,l)} \right) \oint_{S_{k+1}} [\vec{f}_{\tau,s}^{(k+1)} \vec{\mathbb{H}}_{q_0}^{(k)*}] d\vec{S} \right), \quad (1.23)$$

The coupling coefficients $\alpha_{q_0}^{(k,l)}$ depend on geometry of the waveguide under consideration and the frequency $\omega$.

### 1.1 Procedure for calculation of the coupling coefficients on the base of the one eigen mode

In the new approach we also divide the waveguide into the set of arbitrary volumes $V_k$ as shown in Fig. 1 and represent electromagnetic field in $V_k$ as

$$\vec{E} = \vec{\tilde{E}} + e_{q_0}^{(k)} \vec{\mathbb{E}}_{q_0}^{(k)}, \quad \vec{r} \in V_k \quad (1.24)$$

$$\vec{H} = \vec{\tilde{H}} + \frac{\omega}{\omega_{q_0}^{(k)*}} \frac{\varepsilon}{|\varepsilon|} e_{q_0}^{(k)} \vec{\mathbb{H}}_{q_0}^{(k)}, \quad \vec{r} \in V_k \quad (1.25)$$

We will also suppose that amplitudes of the basis mode $e_{q_0}^{(k)}$ can be expressed through the tangential components of the electric fields on the surfaces $S_k$ and $S_{k+1}$

$$\left(\omega_q^{(k)2} - \omega^2\right) e_{q_0}^{(k)} = \frac{i\omega_{q_0}^{(k)*}}{N_{q_0}^{(k)}} \frac{|\varepsilon|}{\varepsilon} \left( \oint_{S_k} [\vec{E}_\tau^{(k)} \vec{\mathbb{H}}_{q_0}^{(k)*}] d\vec{S} + \oint_{S_{k+1}} [\vec{E}_\tau^{(k+1)} \vec{\mathbb{H}}_{q_0}^{(k)*}] d\vec{S} \right), \qquad (1.26)$$

Substituting (1.24) and (1.25) into the Maxwell equations (1.1), we obtain

$$\nabla \times \vec{\tilde{E}} - i\omega\mu_0 \vec{\tilde{H}} = i\mu_0 \frac{\omega^2 - \omega_{q_0}^{(k)2}}{\omega_{q_0}^{(k)*}} \frac{\varepsilon}{|\varepsilon|} e_{q_0}^{(k)} \vec{\mathbb{H}}_{q_0}^{(k)} = -\vec{j}_{mag}^{(k)}, \qquad \vec{r} \in V_k,$$

$$\nabla \times \vec{\tilde{H}} + i\omega\varepsilon_0\varepsilon\vec{\tilde{E}} = 0 \qquad (1.27)$$

Making use of (1.26), we have

$$\nabla \times \vec{\tilde{E}} - i\omega\mu_0\vec{\tilde{H}} = \frac{\mu_0}{N_{q_0}^{(k)}} \left( \oint_{S_k} [\vec{\tilde{E}}_\tau^{(k)} \vec{\mathbb{H}}_{q_0}^{(k)*}] d\vec{S} + \oint_{S_{k+1}} [\vec{\tilde{E}}_\tau^{(k+1)} \vec{\mathbb{H}}_{q_0}^{(k)*}] d\vec{S} \right) \vec{\mathbb{H}}_{q_0}^{(k)} = -\vec{j}_{mag}^{(k)},$$

$$\nabla \times \vec{\tilde{H}} + i\omega\varepsilon_0\vec{\tilde{E}} = 0, \ \vec{r} \in V_k. \qquad (1.28)$$

From (1.28) it f0llows that the fields $\vec{\tilde{E}}, \vec{\tilde{H}}$ satisfy such conditions

$$\int_{V_k} \vec{\tilde{E}}^{(l)} \vec{\mathbb{E}}_{q_0}^{(k)} dV = 0 \qquad (1.29)$$

$$\int_{V_k} \vec{\tilde{H}}^{(l)} \vec{\mathbb{H}}_{q_0}^{(k)} dV = 0 \qquad (1.30)$$

Indeed, if we decompose the fields $\vec{\tilde{E}}, \vec{\tilde{H}}$ in terms of eigen modes

$$\vec{\tilde{E}} = \sum_q \tilde{e}_q^{(k)} \vec{\mathbb{E}}_q^{(k)}, \ z_k < z < z_{k+1}, \qquad (1.31)$$

$$\vec{\tilde{H}} = i\sum_q \tilde{h}_q^{(k)} \vec{\mathbb{H}}_q^{(k)}(\vec{r}), \ z_k < z < z_{k+1}, \qquad (1.32)$$

then we obtain

$$\frac{\varepsilon}{|\varepsilon|} \frac{\left(\omega_q^{(s)2} - \omega^2\right) N_q^{(s)}}{i\omega_q^{(s)*}} \tilde{e}_q^{(s)} = \left( \oint_{S_k} [\vec{\tilde{E}}_\tau^{(k)} \vec{\mathbb{H}}_q^{(k)*}] d\vec{S} + \oint_{S_{k+1}} [\vec{\tilde{E}}_\tau^{(k+1)} \vec{\mathbb{H}}_q^{(k)*}] d\vec{S} \right) + \int_V \vec{j}_{mag}^{(k)} \vec{\mathbb{H}}_q^{(s)*} dV, \ (1.33)$$

The right hand part of (1.33) we can rewrite as

$$\left( \oint_{S_k} [\vec{\tilde{E}}_\tau^{(k)} \vec{\mathbb{H}}_q^{(k)*}] d\vec{S} + \oint_{S_{k+1}} [\vec{\tilde{E}}_\tau^{(k+1)} \vec{\mathbb{H}}_q^{(k)*}] d\vec{S} \right) + \int_V \vec{j}_{mag}^{(k)} \vec{\mathbb{H}}_q^{(s)*} dV =$$

$$= \begin{cases} \left( \oint_{S_k} [\vec{\tilde{E}}_\tau^{(k)} \vec{\mathbb{H}}_q^{(k)*}] d\vec{S} + \oint_{S_{k+1}} [\vec{\tilde{E}}_\tau^{(k+1)} \vec{\mathbb{H}}_q^{(k)*}] d\vec{S} \right), & q \neq q_0 \\ 0, & q = q_0 \end{cases} \qquad (1.34)$$

Therefore, there is no terms with $q = q_0$ in the sums (1.31),(1.32), from which it follows that conditions (1.29) and (1.30) are fulfilled.

$\vec{\mathbb{E}}_{q_0,\tau}^{(k)} = 0$ at the volume borders, then $\vec{\tilde{E}}_\tau$ is a continues vector field throughout the waveguide. $\tilde{E}_z$ and $\vec{\tilde{H}}$ are not a continues ones at the borders between $V_{k-1}, V_k$.

At the surfaces $S_k$ and $S_{k+1}$ (that are the borders between $V_{k-1}, V_k$ and $V_k, V_{k+1}$) the continuity conditions on $\vec{H}$ may be stated as

$$\vec{\tilde{H}}\left(z=z_{k}-0\right)+\frac{\omega}{\omega_{q_{0}}^{(k-1)*}}\frac{\varepsilon}{|\varepsilon|}e_{q_{0}}^{(k-1)}\vec{\mathbb{H}}_{q_{0}}^{(k-1)}\left(z=z_{k}\right)=\vec{\tilde{H}}\left(z=z_{k}+0\right)+\frac{\omega}{\omega_{q_{0}}^{(k)*}}\frac{\varepsilon}{|\varepsilon|}e_{q_{0}}^{(k)}\vec{\mathbb{H}}_{q_{0}}^{(k)}\left(z=z_{k}\right),$$

$$\vec{\tilde{H}}\left(z=z_{k+1}-0\right)+\frac{\omega}{\omega_{q_{0}}^{(k)*}}\frac{\varepsilon}{|\varepsilon|}e_{q_{0}}^{(k)}\vec{\mathbb{H}}_{q_{0}}^{(k)}\left(z=z_{k+1}\right)=\vec{\tilde{H}}\left(z=z_{k+1}+0\right)+\frac{\omega}{\omega_{q_{0}}^{(k+1)*}}\frac{\varepsilon}{|\varepsilon|}e_{q_{0}}^{(k+1)}\vec{\mathbb{H}}_{q_{0}}^{(k+1)}\left(z=z_{k+1}\right),$$

(1.35)

These conditions are linear with respect to the amplitudes of the basis modes $e_{q_0}^{(k)}$, so we can represent the fields $\vec{\tilde{E}}$, $\vec{\tilde{H}}$ as

$$\vec{\tilde{E}} = \sum_{l=-\infty}^{l=\infty} e_{q_0}^{(l)} \vec{\tilde{E}}^{(l)},$$
$$\vec{\tilde{H}} = \sum_{l=-\infty}^{l=\infty} e_{q_0}^{(l)} \vec{\tilde{H}}^{(l)}$$

(1.36)

The fields $\vec{\tilde{E}}^{(l)}$, $\vec{\tilde{H}}^{(l)}$ are the solutions of such equations

$$\nabla \times \vec{\tilde{E}}^{(l)} - i\omega\mu_0 \vec{\tilde{H}}^{(l)} = \frac{\mu_0}{N_{q_0}^{(k)}}\left(\oint_{S_k}[\vec{\tilde{E}}_\tau^{(k,l)} \vec{\mathbb{H}}_{q_0}^{(k)*}]d\vec{S} + \oint_{S_{k+1}}[\vec{\tilde{E}}_\tau^{(k+1,l)} \vec{\mathbb{H}}_{q_0}^{(k)*}]d\vec{S}\right)\vec{\mathbb{H}}_{q_0}^{(k)} \quad \vec{r} \in V_k,$$

(1.37)

$$\nabla \times \vec{\tilde{H}}^{(l)} + i\omega\varepsilon_0 \vec{\tilde{E}}^{(l)} = 0$$

with such conditions at the surfaces $S_k$ and $S_{k+1}$

$$\vec{\tilde{H}}^{(l)}\left(z=z_k-0\right) = \vec{\tilde{H}}^{(l)}\left(z=z_k+0\right) + \frac{\omega}{\omega_{q_0}^{(k)*}}\frac{\varepsilon}{|\varepsilon|}\delta_{k,l}\vec{\mathbb{H}}_{q_0}^{(k)}\left(z=z_k\right),$$

$$\vec{\tilde{H}}^{(l)}\left(z=z_{k+1}-0\right) + \frac{\omega}{\omega_{q_0}^{(k)*}}\frac{\varepsilon}{|\varepsilon|}\delta_{k,l}\vec{\mathbb{H}}_{q_0}^{(k)}\left(z=z_{k+1}\right) = \vec{\tilde{H}}^{(l)}\left(z=z_{k+1}+0\right),$$

(1.38)

where $\delta_{k,l}$ is the Kronecker symbol.

Equations for the amplitudes of the basis mode take then the form

$$\left(\omega_{q_0}^{(k)2} - \omega^2\right)e_{q_0}^{(k)} = \omega_{q_0}^{(k)2}\sum_l e_{q_0}^{(l)}\alpha_{q_0}^{(k,l)},$$

(1.39)

where the coupling coefficients $\alpha_{q_0}^{(k,l)}$

$$\alpha_{q_0}^{(k,l)} = \frac{i\omega_{q_0}^{(k)*}}{\omega_{q_0}^{(k)2}N_{q_0}^{(k)}}\frac{|\varepsilon|}{\varepsilon}\left(\oint_{S_k}[\vec{\tilde{E}}_\tau^{(k,l)} \vec{\mathbb{H}}_{q_0}^{(k)*}]d\vec{S} + \oint_{S_{k+1}}[\vec{\tilde{E}}_\tau^{(k+1,l)} \vec{\mathbb{H}}_{q_0}^{(k)*}]d\vec{S}\right)$$

(1.40)

If we find vector functions $\vec{\tilde{E}}_\tau^{(k,l)}$ that are the solution of equations (1.37) and fulfilled the continuity conditions (1.38), we can calculate the coupling coefficients $\alpha_{q_0}^{(k,l)}$.

### Conclusions

We present the new method for calculation of the coupling coefficients in the Coupling Cavity Model of arbitrary chain of resonators that used only one eigen vector. The detailed description of application this method to 1-D geometry will be given in the next paper.